\documentclass[
    ,final            % use final for the camera ready runs
  ]
  {aipproc}

\layoutstyle{6x9}

\begin{document}

\title{Single-Spin Beam Asymmetry in Semi-Exclusive Deep-Inelastic Electroproduction}

\author{Andrei Afanasev}{ 
address={Jefferson Lab, 12000 Jefferson Ave., Newport News, VA 23606, USA}
}

\author{C. E. Carlson$^*$}{
 address={Department of Physics, College of William and Mary, 
  Williamsburg, VA 23187, USA} 
}

\begin{abstract}
Recent measurements from Jefferson Lab show significant beam single 
spin asymmetries in deep inelastic scattering.
 The asymmetry is due to interference 
of  longitudinal and transverse photoabsorption amplitudes which 
have different phases induced by the final-state interaction between  
the struck quark and the target spectators. We developed a dynamical model 
for a single-spin beam asymmetry in deep-inelastic scattering. 
Our results are consistent  with the experimentally observed magnitude of 
this effect. We conclude that similar mechanisms involving quark orbital 
angular momentum (`Sivers effect') are responsible for both target 
and beam single-spin asymmetries. 
\end{abstract}

\maketitle

%%%%%%%%%%%%%%%%%%%%%%%%%%%%%%%%%%%%%%%%%%%%
%% MAINMATTER
%%%%%%%%%%%%%%%%%%%%%%%%%%%%%%%%%%%%%%%%%%%%

Single--Spin Asymmetries (SSA) in semi--exclusive electroproduction
reactions provide important new information about the nucleon spin structure
\cite{sivers91,collins92,jjmt}. Recent successful measurements of SSA
in leptoproduction \cite{hermes_ssa,CLAS02} and anticipated experiments
at Jefferson Lab after the 12-GeV energy upgrate stimulate a lot of interest in this
subject.

One can relate SSA to the products of parton distribution and fragmentation
functions, see Ref.\cite{baronereview} for review. In the quark--parton
model, the origins of SSA can be attributed both to the time--reversal odd quark 
distribution, often called the `Sivers effect' \cite{sivers91}, and to the 
time--reversal odd fragmentation also known as `Collins effect' \cite{collins92}.
Recently, Brodsky and collaborators \cite{bhs} used a model to demonstate that
the target SSA can be generated at a leading twist level without
time--reversal odd fragmentation. It led to the revision \cite{collins,bjy} of 
time--reversal arguments, validating phenomenology with time--reversal odd 
parton distributions. 

It has been known for a long time that single--spin polarization observables
in particle scattering are caused by the spin--orbit interaction
\cite{mott,nucl}. 
SSA in deep--inelastic scattering should become a sensitive probe of the orbital
angular momentum (OAM) of the participating quarks since SSA would simply
vanish without contributions from both spin and OAM.  It was demonstrated 
\cite{bhs} that indeed OAM of quarks is crucial for describing target SSA.

In this work, we extend the calculation of SSA to the case of a polarized electron
beam using the model  
 that Ref.\cite{bhs} used for the polarized target case.
The model is reminiscent of model decomposition of the proton studied earlier in 
Ref.\cite{ezawa-drell-lee}. We demonstrate that a mechanism similar to Ref.\cite{bhs} 
is also generating beam SSA of a magnitude compatible with
experimental data. The considered mechanism does not require the Collins fragmentation
function and therefore does not allow relation of the measured asymmetry to
the quark distribution $e(x)$ introduced by Jaffe and Ji \cite{jjmt}.
In this respect, our result for the beam SSA differs from analyses of
Refs.\cite{leveltmulders,efremovgoeke}.

\begin{figure}[ht]
  \includegraphics[height=.2\textheight]{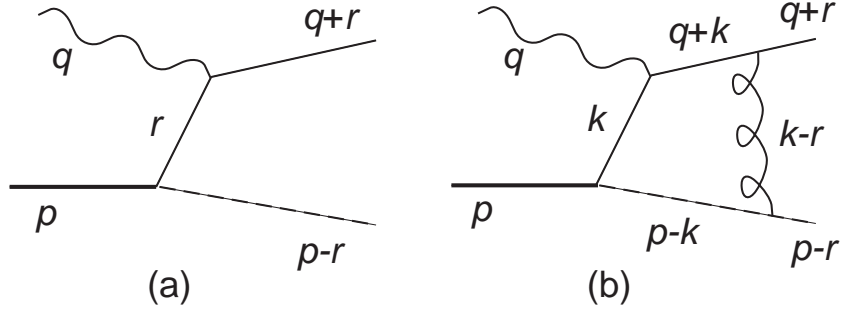}
  \caption{(a) Tree--level diagram and (b) Final--state gluon exchange.}
\label{feyndiag}  
\end{figure}

We calculate the beam SSA $A_{LU}^{\sin\phi}$ measured in the experiment and defined as
\begin{equation}
d\sigma\propto(1+h_e A_{LU}^{\sin\phi} \sin\phi),
\end{equation}
where $h_e=\pm 1$ indicates the electron beam helicity, and $\phi$ is the angle between
the planes of electron scattering and hadron production.
Following Ref.\cite{bhs}, we describe the deep--inelastic electroproduction process 
with a model of a spin 1/2 proton of mass $M$ which is made from charged spin--1/2 and
spin--0 constituents of masses $m$ and $\lambda$, respectively (Fig.\ref{feyndiag}).
During the scattering, the virtual photon is absorbed by an active spin--1/2 parton,
while the scalar diquark is used to describe the spectator system. The momentum of the
knocked--out quark is set equal to the observed hadronic momentum.
The proton--quark--diquark vertex function should be a solution to a bound--state equation,
which we here simplify as a scalar vertex with a coupling parameter $g$.

The leading--order diagram Fig.\ref{feyndiag}a alone produces zero asymmetry 
$A_{LU}^{\sin\phi}$, since its contribution to the scattering amplitude is purely real. 
The one--loop next--to--leading--order correction Fig.\ref{feyndiag}b needs to be
included in order to generate nonzero beam and target SSA. 
The loop diagram Fig.\ref{feyndiag}b contribution is given by the formula
\begin{equation}
J_{\mu}^{NLO}=i e_q g C_F 4 \pi \alpha_s \int \frac{d^4k}{(2 \pi)^4} 
\frac{{\bar u}(q+r)\gamma_{\nu} ({\not\! q}+{\not\! k}+m)\gamma_{\mu} ({\not\! k}+m)(-g_{\nu \tau})(2
p-k-r)_{\tau}}{D_1 D_2 D_3 D_4},
\label{loop}
\end{equation}
 where $C_F=4/3$ and $D_i$ are denominators that come from the four propagators.
 
 We calculate the contribution of the Fig.\ref{feyndiag} mechanism to the anti--symmetric
 part of the hadronic tensor 
defined by the imaginary part of the intereference between Fig.\ref{feyndiag}a and Fig.\ref{feyndiag}b
amplitudes. Cutkosky rules are used to calculate the absorptive part of the loop contribution, 
replacing two of the denominators $D_i$ in Eq.(2) by the corresponding delta functions as follows:
\begin{equation}
\frac{1}{((q+k)^2-m^2+i\epsilon)((p-k)^2-\lambda^2+i\epsilon)}\to (2 \pi i)^2 \delta((q+k)^2-m^2) \delta((p-k)^2-\lambda^2),
\end{equation} 
where $m$ and $\lambda$ are the masses of the quark and diquark, respectively.
As a result, the 4--dimensional integration over the loop momentum in Eq.(2) is reduced to
a 2--dimensional angular integration, making the result safe from ultraviolet divergence.
The beam asymmetry is generated due to interference between absorption of longitudinal
and transverse virtual photons (LT-interference) which acquire different phases as a 
result of final--state gluon exchange. In the scaling limit, the asymmetry is proportional
to $r_T/\sqrt{Q^2}$, where $r_T$ is a transverse component of the final quark momentum.
The factor $1/\sqrt{Q^2}$ points to the twist--three origin of this observable.
The calculations were also performed in the light--cone gauge, producing the same results. 

The amplitude Fig.\ref{feyndiag}b is infrared--divergent as the momentum of the exchanged
gluon approaches zero. However, the infrared--divergent terms exactly cancel at the level
of observable asymmetry. The same spin--dependent mechanism reduces relative 
contribution of soft gluons to the integral Eq.(\ref{loop}).
We also note that we neglected the loop correction to the unpolarized cross
section, otherwise we would have to add real gluon bremsstrahlung or introduce an 
infared cut--off to make the theory infrared--safe. 
\begin{figure}[t]
  \includegraphics[height=.4\textheight]{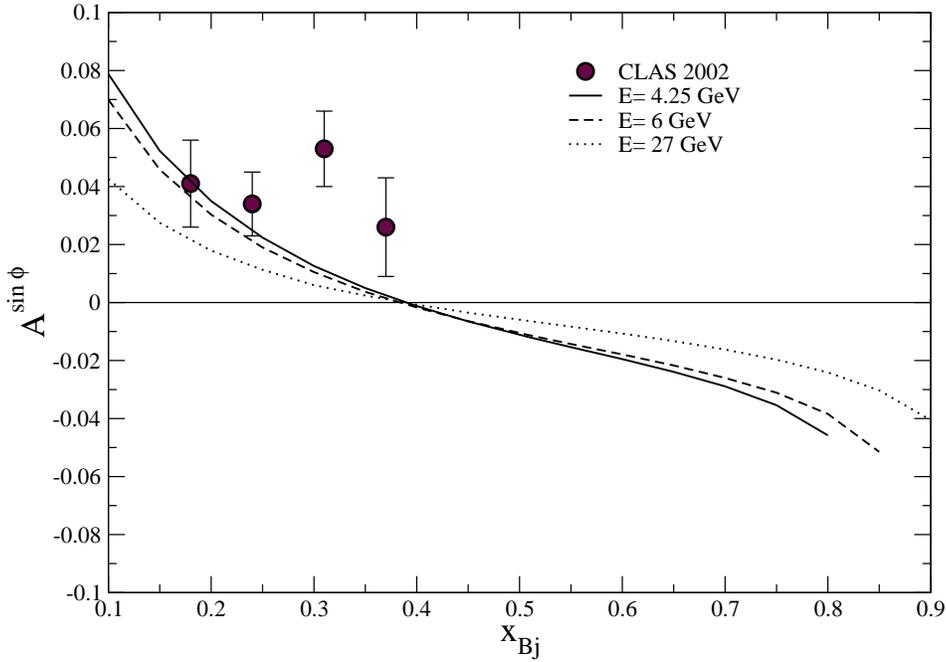}
  \caption{Beam SSA as a function of $x_{Bj}$ for different beam energies at fixed $y$= 0.7 
  and $r_T$= 0.4 GeV/c. Data points are from Ref.\cite{CLAS02} for 
  the process $p(\vec e,e'\pi^+)X$. }
  \label{asym}
\end{figure}

The results of our model for  beam SSA is shown in Fig.\ref{asym} for different beam energies as
a function of Bjorken $x_{Bj}$.
We used the values of the parameters $m=$ 0.3 GeV, $\lambda$= 0.8 GeV and $\alpha_s$= 0.3.
The value of the kinematic variable $y=$ 0.7 and beam energies of 4.25 and 6 GeV are chosen 
to match the experimental conditions \cite{CLAS02}. One can see that the prediction for beam SSA is
consistent with its observed magnitude. For higher beam energies of 27 GeV relevant to the 
HERMES experiment \cite{hermes_ssa} and fixed values
of the kinematic variables $x_{Bj},y$, the beam asymmetry is supressed by about a factor of
two, following the $1/\sqrt{Q^2}$ dependence. 
Our model predicts that the asymmetry changes its
sign around $x=$ 0.5, where no experimental data are available. Thus extension of experimental measurements to higher 
values of $x$ can test validity of the proposed mechanism. Note that
beam SSA calculated in a simplified model with scalar partons is approximately constant as a function of $x_{Bj}$
and equals about +0.08 for the 6-GeV kinematics of Fig.\ref{asym}.

Let us discuss the relation of the obtained results to the target SSA calculations of Ref.\cite{bhs}.
Target SSA are due to the T-odd observable $\vec S_p \cdot \vec q \times \vec p_h$ made from the 
proton polarization vector $\vec S_p$, the photon momentum $\vec q$, and the outgoing hadron's momentum $\vec p_h$.
The essential ingredient of the model \cite{bhs} is quark OAM in the initial nucleon state,
so that the struck parton may have helicities both parallel and antiparallel to the nucleon helicity.
Interference between transitions to different helicity states, supplemented by the phase difference
due to final gluon exchange is the mechanism that generates the target SSA. Very similarly,
beam SSA is due to the correlation  $\vec S_\gamma \cdot \vec q \times \vec p_h$, where  
$\vec S_\gamma$ is the virtual photon polarization. The beam SSA arises due to intereference
between longitudinal and transverse photoabsorption amplitudes, with the transverse momentum and
phase differences generated by the same gluon--exchange mechanism as in the polarized target case.
The quark OAM now unambiguosly enters at the photon--parton level and the result does not depend on
the quark OAM contribution to the nucleon light--cone wave function.  Therefore, measurements of beam SSA
are needed for adequate interpretation of target SSA measurements within a parton model.
By analogy with the target asymmetry and in absence of T--odd fragmentation,
 we can view the considered mechanism as a 
{\it photon Sivers effect}, 
emphasizing the importance of this effect in polarized deep--inelastic scattering.

 We acknowledge useful discussions with H. Avakian, S. Brodsky and D.S. Hwang.
 This work was supported by the US Department of Energy under contract 
 DE-AC05-84ER40150. In addition, C.E.C. thanks NSF for support under grant
 PHY-0245056.

%% If the bibliography is produced without BibTeX comment out the

\end{document}